\documentclass[a4paper]{article}
\usepackage{latexsym}
\begin{document}

\newtheorem{theorem}{Theorem}
\newtheorem{proposition}{Proposition}
\newtheorem{remark}{Remark}
\newtheorem{corollary}{Corollary}
\newtheorem{lemma}{Lemma}
\newtheorem{observation}{Observation}
\newtheorem{fact}{Fact}

\newcommand{\qed}{\hfill$\Box$\medskip}

\title{The Power of Centralized PC Systems of Pushdown Automata\\
       {(Preliminary Report)}} 
\author{Holger Petersen\\
Reinsburgstr. 75\\
70197 Stuttgart\\
Germany} 

\maketitle

\begin{abstract}
Parallel communicating systems of pushdown automata (PCPA) were introduced in (Csuhaj-Varj{\'u} et.\ al.\ 2000) and 
in their centralized variants shown to
be able to simulate nondeterministic one-way multi-head pushdown automata.
A claimed converse simulation for returning mode (Balan 2009) turned out to be incomplete (Otto 2012) and a language 
was suggested for separating these PCPA  of degree two (number of pushdown automata)
from nondeterministic one-way two-head pushdown automata.
We show that the suggested language can be accepted by the latter computational model. 
We present a different example over a single letter alphabet 
indeed ruling out the possibility of a 
simulation between the models.
The open question about the power of centralized PCPA working in returning mode is then settled by showing them to
be universal. Since the construction is possible using systems of degree two, this also improves the previous bound three for generating all 
recursively enumerable languages.
Finally PCPAs are restricted in such a way that a simulation by multi-head automata is possible. 
\end{abstract}

\section{Introduction}  
Parallel communicating systems of pushdown automata (PCPA) were introduced in \cite{CMMV00} and some 
properties of such systems were shown. Among these are that general PCPA of degree two (number of components)  and returning PCPA
of degree three can generate all recursively enumerable languages.

In  \cite{Balan09} a proof was presented that centralized PCPA
working in returning mode can be simulated by multi-head pushdown automata. 
More precisely, these PCPA of degree $k$ (with $k$ pushdown automata) were claimed to be simulated 
by nondeterministic one-way $k$-head automata. Since the converse simulation had previously been shown in \cite{CMMV00}, this 
would be an interesting characterization in
contrast to the universal power of other variants of non-centralized or non-returning PCPA 
(see \cite{CMMV00} and \cite{Balan09} for further references). It would also
link the recent investigation of PCPA to classical automata theory dating back more than 40 years \cite{HI68}.

As demonstrated by Otto \cite{Otto12}, the proof from \cite{Balan09} is incomplete and the power of centralized PCPA working 
in returning mode is open. He defined  a language that can be accepted by
centralized PCPA of degree two working in returning mode for which the simulation given in \cite{Balan09} fails.

The main purpose of the present paper is to elaborate on the observations from \cite{Otto12} about the inherent synchronization of PCPAs  
and settle the open problem resulting from the gap in \cite{Balan09}. First we show that the language from  \cite{Otto12} on which the simulation 
fails can be accepted in a  way deviating substantially from the suggested simulation by a 
nondeterministic two-head pushdown automaton (in \cite{Otto12} it is stated without proof that four heads are sufficient for this task). 
Since for degree two this is the claim of \cite{Balan09}, the language cannot serve as
a witness showing the claimed result to be incorrect. In order to obtain a concrete counterexample we show that a non-regular
language over a single-letter alphabet can be accepted by a centralized  PCPA in returning mode. By a classical result such languages 
cannot be accepted by one-way multi-head pushdown automata.

We then present the main result that centralized PCPA of degree two working in returning mode
are universal, placing them among the formally more powerful systems mentioned above. 
It also improves the result that non-centralized PCPA of degree three can accept every 
recursively enumerable language \cite[Theorem~4]{CMMV00} and answers several open problems from Section~5 of
\cite{CMMV00}.
Next we restrict the PCPA to linear time and give a simulation
by multi-head pushdown automata in the spirit of  \cite{Balan09}. This simulation is probably not tight, since it uses more heads than the degree of
the PCPA and makes use of sensing (detecting coincidence of heads).

\section{Preliminaries}
Several variants of  PCPA  were defined in \cite{CMMV00,Balan09}. Informally, a PCPA of degree $k$ consists
of a collection of $k$ nondeterministic pushdown automata in the classical sense. These automata (called components)
work in a synchronous fashion reading the same input string. Note that by epsilon-moves on the input the components can process
the input at different speeds.
Communication is carried out via special pushdown store symbols. If one of these symbols is on top of a pushdown store, 
instead of a usual step the contents of another pushdown store are copied to the pushdown store replacing the topmost symbol. If the PCPA is working in  
returning mode, the source pushdown store is emptied up to a bottom symbol. The PCPA is centralized if only one component (say the first) can
use communication symbols. An input is accepted if all components have read the entire input string and reach final states.

We also give a formal definition of PCPA since we will need it for a concrete example.
A PCPA of degree $k$ is a tuple 
$$ A=(V, \Delta, A_1, A_2,\ldots, A_k, K)$$
where 
\begin{itemize}
\item $V$ is a finite input alphabet,
\item $\Delta$ is a finite alphabet of pushdown symbols,
\item $A_i$ is a component as defined below for $1 \le i \le k$,
\item $K = \{ K_1, \ldots, K_k\}\subseteq \Delta$ is a set of query symbols.
\end{itemize}

Each component $A_i = (Q_i, V, \Delta, f_i, q_i, Z_i, F_i)$ is a pushdown automaton where
\begin{itemize}
\item $Q_i$ is a finite set of states,
\item $f_i$ is a function from $Q_i\times (V\cup \varepsilon)\times\Delta$ to the finite subsets of  $Q_i\times\Delta^*$,
\item $q_i\in Q_i$ is the initial state,
\item $Z_i\in \Delta$ is the bottom symbol,
\item $F_i\subseteq Q_i$ is the set of final states.
\end{itemize}
If only function $f_1$ of the first component maps to sets with members containing query symbols, the system is called centralized.

A configuration of a PCPA of degree $k$ is a $3k$-tuple
$$(s_1, x_1, \alpha_1, \ldots, s_k, x_k, \alpha_k)$$
where
\begin{itemize}
\item $s_i\in Q_i$ is the state of component $A_i$,
\item $x_i\in V^*$ is the part of the input not yet processed by $A_i$,
\item $\alpha_i\in \Delta^*$ is the word on the pushdown store of $A_i$ with its topmost symbol on the left.
\end{itemize}
In returning mode the step relation $\vdash_r$ between configurations is defined by:
$$(s_1, x_1, B_1\alpha_1, \ldots, s_k, x_k, B_k\alpha_k) \vdash_r (s_1', x_1', \alpha_1', \ldots, s_k', x_k', \alpha_k'),$$
if one of the following conditions holds:
\begin{description}
\item[Internal step:] $\{B_1, \ldots, B_k\}\cap K=\emptyset$, $x_i = a_ix_i'$ with $a_i \in V \cup\{\varepsilon\}$, 
  $(s_i', \beta)\in f_i(s_i, a_i, B_i)$ with $\alpha_i' = \beta\alpha_i$.
\item[Communication step:] $\{B_1, \ldots, B_k\}\cap K\neq\emptyset$, for each $B_i = K_{j_i}$ with $B_{j_i}\not\in K$ we have
  $\alpha_i' =  B_{j_i}\alpha_{j_i}\alpha_i$, $\alpha_{j_i}'=Z_{j_i}$, and $\alpha_m' = B_m\alpha_m$ for all other indices $m$.
  States and input are not modified: $s_i' =  s_i$ and $x_i' =  x_i$ for $1 \le i \le k$.
\end{description}
The PCPA accepts exactly those words $w$ that admit a sequence of steps from the initial configuration
$$(q_1, w, Z_1, \ldots, q_k, w, Z_k)$$
to a final configuration
$$(s_1, \varepsilon, \alpha_1, \ldots, s_k, \varepsilon, \alpha_k)$$
with $s_i\in F_i$ for $1 \le i \le k$.

A nondeterministic one-way multi-head pushdown automaton has $k$ read-only heads it can move forward 
on a single input tape and one pushdown store. In some constructions we assume that the input-heads are sensing 
(they can ``see'' each others when they scan the same position). This is a variant of the model investigated in \cite{CL88}.
Formal definitions of  these automata can be found in \cite{HI68,WW86,CMMV00}. 
In the first reference input-tapes have end-markers and we also assume this here.
When we mention multi-head pushdown automata we always refer to the one-way variant in the present paper.

\section{Results for General Centralized PCPA}
We start our investigation with the language defined in \cite{Otto12} and show that it can be accepted by the type of automata
proposed in \cite{Balan09} for a simulation of centralized PCPA working in returning mode. Notice that this does not mean that 
the simulation works, since our algorithm deviates from the simulation of \cite{Balan09} as outlined for this specific language in \cite{Otto12}. It can however be deduced that
a different witness language is necessary to show that the claim of \cite{Balan09} in general fails.
\begin{theorem}
Language $L = \{ uvu^Rv^Ru^R \mid u, v\in \{a, b\}^{+}, |u| = |v|\}$ (where $w^R$ is the reversal of string $w$) from \cite{Otto12} can be accepted by a 
nondeterministic two-head pushdown automaton.
\end{theorem}
Proof.
We first describe the algorithm carried out by a nondeterministic two-head pushdown automaton.
\begin{enumerate}
\item While moving head~1 forward, push a non-empty prefix $x$ of the input $w$ onto the pushdown store. The head stops at a nondeterministically chosen position.
\item Move head~2 over the input counting the number of steps using a special counting symbol. Head~2 stops at a nondeterministically chosen position.
\item Move head~1 and head~2 in parallel over the input at least one position and compare the symbols read. Also pop four counting symbols for each step and repeat until 
   all counting symbols have been removed. 
  Reject the input in one of the following cases:
  \begin{itemize}
  \item A head reaches the end-marker before all counting symbols have been removed.
  \item The symbols read by head~1 and head~2 are not equal.
  \item The remaining number of counting symbols in an iteration is between 1 and 3 (which means that the initial number was not divisible by 4).
  \item Head~2 has not reached the end-marker when all counting symbols have been removed.
  \end{itemize}
\item With the help of head~1 compare the contents of the pushdown store to the remaining part of the input.
\item Check that the end-marker is reached exactly when the pushdown store becomes empty and accept.
\end{enumerate}

Clearly every string in $L$ can be accepted by the algorithm described above.
On arbitrary input $w$ with $|w| = n$, a prefix $x$ of $w$ is first pushed onto the pushdown store. Then a suffix $y$ of the input is compared to a
section of equal length after $x$. By counting on the pushdown store the automaton ensures
$$ |y| = (|w|-|y|)/4$$
and thus
$$|y| = |w|/5.$$
Suppose that the copies of $y$ being compared overlap. Then $|x| > |w| - 2|y| = 3|w|/5 \ge |w|/2$ and the last comparison will fail. Therefore
we can assume that $w = xyzy$ and in case of acceptance $x^R = zy$. It follows that $|x| = 2|w|/5$, $|z| = |w|/5$ and by letting $u = y^R$ and $v=z^R$ 
we see that the input 
$$w =  xyzy = (zy)^Ryzy = y^Rz^Ryzy = uvu^Rv^Ru^R$$
belongs to $L$.
\qed

Next we present a non-regular language accepted by a centralized PCPA of degree two working in returning mode. This improves the construction 
from \cite[Example~1]{CMMV00} 
where the same language was shown to be accepted by a similar non-centralized system of degree four. At the same time it will serve as one
building-block of a proof that the main claim from \cite{Balan09} is incorrect.

{\bf Example:} The language $\{ a^{2^n} \mid n\ge 1\}$ can be generated by a centralized PCPA of degree 2 working in returning mode
with the following transitions of its components:
\begin{eqnarray*} 
f_1(q_0^1, a, Z_1) & = & \{ (q_1^1, Z_1)\}\\
f_1(q_0^1, a, a) & = & \{ (q_1^1, a)\}\\
f_1(q_1^1, a, Z_1) & = & \{ (q_2^1, K_2)\}\\
f_1(q_1^1, a, a) & = & \{ (q_1^1, \varepsilon)\}\\
f_1(q_2^1, a, Z_1) & = & \{ (q_1^1, K_2)\}\\
f_1(q_2^1, a, a) & = & \{ (q_1^1, \varepsilon)\}\\
f_1(q_0^2, a, Z_1) & = & \{ (q_1^1, Z_1)\}\\
f_1(q_0^2, a, a) & = & \{ (q_1^1, a)\}\\
f_2(q_1^2, a, Z_2) & = & \{ (q_1^2, aZ_2)\}\\
f_2(q_1^2, a, a) & = & \{ (q_1^2, aaa)\}
\end{eqnarray*} 
Initial states are $q_0^1$ and $q_0^2$, final states are $F_1 = \{ q_2^1 \}$ and $F_2 = \{ q_1^2 \}$.

We outline the idea of the construction. First both components read a single $a$. Then component~1 reads as many input symbols as
indicated by the size of its pushdown store. Component~2  in  parallel pushes approximately twice as many symbols, with the exception of the first
step since the bottom symbol is already present. This process ends when the pushdown store of component~1 is empty. Then the contents of
the pushdown store of component~2 is copied and the process is repeated.

Component~2 stays in its final state after the first step. Therefore acceptance depends on component~1, which is in a final state after the 
pushdown store has become empty. This happens after reading
$$1 + \sum_{i=0}^{k}2^i = 2^{k+1}$$
input symbols for some $k\ge 0$.

We consider the computation on $a^8$:
\begin{eqnarray*} 
 (q_0^1, a^8, Z_1, q_0^2, a^8, Z_2) & \vdash_r & (q_1^1, a^7, Z_1, q_1^2, a^7, Z_2)\,\vdash_r \\
 (q_2^1, a^6, K_2, q_1^2, a^6, aZ_2) & \vdash_r & (q_2^1, a^6, aZ_1, q_1^2, a^6, Z_2) \,\vdash_r \\
 (q_1^1, a^5, Z_1, q_1^2, a^5, aZ_2) & \vdash_r & (q_2^1, a^4, K_2, q_1^2, a^4, aaaZ_2) \,\vdash_r \\
 (q_2^1, a^4, aaaZ_1, q_1^2, a^4, Z_2) & \vdash_r & (q_1^1, a^3, aaZ_1, q_1^2, a^3, aZ_2)\, \vdash_r \\
 (q_1^1, a^2, aZ_1, q_1^2, a^2, aaaZ_2) & \vdash_r & (q_1^1, a, Z_1, q_1^2, a, aaaaaZ_2)\, \vdash_r \\
 (q_2^1, \varepsilon, K_2, q_1^2, \varepsilon, a^7Z_2)
\end{eqnarray*} 
Since in the last configuration all components have read the entire input and all states are final, the word $a^8$ is accepted.

The following result \cite[Theorem~4.2]{HI68} shows a limitation of pushdown automata:
\begin{fact}\label{slamulthead}
The  single-letter alphabet languages accepted by  nondeterministic  multi-head pushdown automata are 
the regular languages over single-letter alphabets.
\end{fact}
The more general result that bounded languages accepted by bounded-reversal multi-head pushdown automata are semilinear 
has been shown in \cite{Ibarra74}.

Notice that a statement analogous to Fact~\ref{slamulthead} about automata with sensing heads is not true. The above language can be accepted
by a deterministic  pushdown automaton with two sensing heads that keeps doubling the distance between the heads
using its pushdown store as a counter.

The language in the example above is not regular, therefore we obtain:
\begin{observation}\label{slasep}
No general simulation of centralized PCPA of degree two working in returning mode by nondeterministic  multi-head automata
is possible.
\end{observation}
We can also conclude that Theorem~8 of \cite{Balan09} (which is weaker than his Theorem~5) 
is wrong, since the system in the example above is simple.
A simple system as defined in \cite{Balan09} has no two components querying the same component and components that query do not
communicate to any other querying component. For a simple centralized PCPA $A$ working in returning mode Balan claimed in his 
Theorem~8 that the accepted language $L(A)$ could be written as
$$L(A) = L(M_1)\cap L(M_2)\cap \cdots\cap L(M_m),$$
where every $M_i$ is a multi-head automaton. Since regular languages are closed under intersection and over a single letter
alphabet each  $L(M_i)$ is regular by Fact~\ref{slamulthead}, this claim contradicts the example (we could also argue that
languages accepted by  multi-head automata are closed under intersection, where the number of heads required for the intersection 
is at most the sum of the heads for the  languages in the intersection).

After the basic separation of the models claimed to be equivalent in \cite{Balan09}, it remains to be investigated how
powerful centralized PCPA working in returning mode really are. The following
result gives an answer and improves the previous bound three for non-centralized systems \cite{CMMV00}.
\begin{theorem}
Every recursively enumerable language can be accepted by a centralized PCPA of degree two working in 
 returning mode.
\end{theorem}
Proof. One register machines with the operations multiplication and conditional division by 2 or 3 are universal, 
see Theorem~14.2-1 of \cite{Minsky67}. We will outline a simulation of such a machine with input tape by a 
PC system of two pushdown automata. 

Component 1 carries out the main task of the simulation. For every instruction of the register machine, an in general unbounded number steps
of the pushdown automaton will be carried out. On its pushdown store a counter is simulated.

Component 2 constantly works in a cycle of length 6, pushing a single counting symbol in every cycle. If the top-most symbol on the pushdown store is the bottom symbol, it is kept and the first
counting symbol is pushed after six steps.Then the cyclic behaviour starts. 

We will first outline how a reset to an empty pushdown store of component~2 can be enforced by component~1. The difficulty is that a communication step transferring the contents of the pushdown store
of component~2 to component~1 empties the pushdown store of component~2 but possibly puts ``garbage'' generated in previous steps on top of the pushdown store of component~1. 
Therefore component~1 repeatedly carries out the following process:
\begin{itemize}
\item A communication step from component~2 to component~1. 
\item Component~1 checks whether the top-most symbol of its pushdown store is the bottom symbol transferred from component~2. If so, the symbol is removed and the process is terminated.
\item Otherwise a loop of component~1 starts that removes one symbol from the pushdown store in each iteration, stopping after the bottom symbol transferred from component~2 has been removed.
\end{itemize}

Note that each execution of the process reduces the number of counting symbols on the pushdown store of component~2 by at least a factor of 6. Thus eventually
only the bottom symbol is transferred from component~2 leaving its pushdown store empty and the pushdown store of component~1 unchanged.

Now we are ready to describe the simulation of register machine instructions by component~1:
\begin{description}
\item[Read a symbol from the input tape:] Carry out transitions reading input without changing the pushdown store.
\item[Multiply by 2 (3):] Reset the pushdown store of component~2. Then carry out a loop of length 12 (18) that removes one counting symbol from 
the pushdown store in each iteration. Finally the bottom symbol is replaced with the communication symbol and the pushdown store is transferred.
\item[Divide by 2 (3):] Reset the pushdown store of component~2. Then carry out a loop of length 3 (2) that removes one counting symbol from 
the pushdown store in each iteration. Determine the remainder by the state when the bottom symbol is read and
replace it with the communication symbol.
\end{description}
If the register machine accepts its input, component~1 enters an accepting state and
we let all states of component~2 be accepting.
\qed

Clearly every language accepted by multi-head pushdown automata is decidable, therefore we obtain another  
separation of PCPA and this computational
model in addition to Observaition~\ref{slasep}. From the known hierarchy results \cite{Chrobak86,CL88} the concrete example $$L' = \{ w_1\#\cdots\#w_n\$ w_n'\#\cdots\#w_1' \mid
w_i\neq w_i'\mbox{ for some $i$}\}$$ can be derived \cite[proof of Theorem~2]{CL88} that separates the classes.

\section{Results for Time-Bounded Centralized PCPA}
In view of the impossibility of simulating centralized PCPA working in returning mode by multi-head pushdown automata we explore in
this section restrictions of PCPA that admit a simulation.

See \cite[Theorem~13.15.8]{WW86} for a proof  that there is no loss of generality in requiring a linear time bound
for  multi-head pushdown automata. Since the
construction from the proof of \cite[Theorem~5]{CMMV00} gives a system accepting in a time proportional to the automaton 
being simulated, we obtain the following:
\begin{observation}\label{simuPDAlin}
Every $k$-head pushdown automaton can be simulated by a centralized PCPA of degree $k$ working in 
returning mode and accepting in linear time.
\end{observation}
This raises the question, whether a result in the spirit of the simulation of \cite{Balan09} is possible when restricting the time bound of
PCPA to being linear. We were not able to give a characterization but present here a partial result that uses additional heads exceeding
the degree of the system. We also assume that the heads are sensing. Observe that such automata are stronger than their counterparts
with non-sensing heads as pointed out after Fact~\ref{slamulthead}.
\begin{theorem}\label{simuPCPAlin}
Every centralized PCPA of degree $k$ working in 
 returning mode and accepting in linear time is simulated by a $2k$-head pushdown automaton with sensing heads.
\end{theorem}
Proof. Let the PCPA accept in at most $cn$ steps on inputs of length $n$ for some constant $c$. 
The heads of the simulator are divided into two groups. 
Heads of the first group simulate the access to the input for the $k$ components of the PCPA. Heads of the second group serve a clocks
measuring the number steps carried out by the components being simulated. 
The multi-head automaton simulates the first component of the PCPA step by step until communication occurs. It also records the
current state of each of the other components. For every step being simulated the
clock for component 1 is advanced by one counting modulo $c$ in the finite control and recording each iteration of this counting process 
by advancing the input head. If communication with component $i > 1$ takes place, the component $i$ is simulated starting at the currents state
using its input head and 
clock until the clock of component 1 and $i$ coincide and the simulation of component 1 continues. 
This can be detected by the sensing property. The pushdown store of component $i$ is simulated 
on top of the contents of the current pushdown store. The first phase of the simulation is terminated if component 1 accepts or if the clock reaches $cn$.
Notice that in the latter case the simulator can reject the input, since by definition all components have to accept. If  component~1 accepts,
the simulator simulates the remaining $k-1$ components in turn until each of them has executed the same number of steps as component~1. 
The input is accepted if  all components accept. 
\qed

When all components of a PCPA read their input synchronously, we can identify heads accessing the input and clocks for each component.
We obtain 
the following result by basically the same simulation as in the proof of Theorem~\ref{simuPCPAlin}:
\begin{theorem}
Every centralized PC system of pushdown automata of degree $k$ working in 
  returning mode without $\varepsilon$-transitions on the input can be simulated by a $k$-head pushdown automaton with sensing heads.
\end{theorem}

\section{Discussion}
We have settled the open problem about the power of centralized PCPA of degree two working in returning mode 
from \cite{CMMV00,Otto12} by showing them to be universal.  This also improves the previous bound three for generating all 
recursively enumerable languages with non-centralized systems and solves several open problems from Section~5 of
\cite{CMMV00}:
\begin{enumerate}
\item PCPA of degree two working in returning mode are as powerful as those of degree three.
\item Centralized systems can accept all recursively enumerable languages.
\item The inclusion of \cite[Theorem~5]{CMMV00} referring to returning mode is strict and becomes a corollary of our main result.
\item The degree hierarchy for centralized PCPA working in returning mode is finite, with the context-free languages at level one and the recursively enumerale languages
 at all other levels.
\end{enumerate}
The simulation of register machines is deterministic, an aspect that is mentioned at the end of Section~5 of
\cite{CMMV00}. In hindsight the power of these systems is surprising, since the claim of \cite{Balan09} would have
implied decidability even in linear time on nondeterminstic Turing machines \cite[Theorem~13.15.8]{WW86}. 

In addition we described
simulations of restricted PCPA by multi-head automata in the spirit of \cite{Balan09}. The optimality of the simulations in terms of 
input heads remains an open question as well as the possibility of a converse simulation, since we required the heads to be sensing. 
We expect that Observation~\ref{simuPDAlin} can be strengthened in this direction.

\section*{Acknowledgement}
Many thanks to Friedrich Otto for remarks on an early draft of this paper.

\end{document}